\documentclass{jfm}

\usepackage{amssymb,amsfonts,amsmath}
\usepackage{graphicx}
\usepackage{natbib}
\usepackage[utf8]{inputenc}

\newtheorem{thm}{Theorem}

\renewcommand{\d}[1]{\ensuremath{\operatorname{d}\!{#1}}}
\newcommand{\D}[1]{\ensuremath{\operatorname{D}\!{#1}}}
\newcommand{\bD}[1]{\ensuremath{\operatorname{\mathbf D}\!{#1}}}
\DeclareMathOperator{\Id}{Id}

\DeclareMathOperator*{\trace}{trace}
\DeclareMathOperator*{\sign}{sign}

\DeclareMathOperator*{\argmax}{arg\,max}
\DeclareMathOperator{\LAVD}{LAVD}

\DeclareMathOperator{\area}{area}

\DeclareMathOperator{\neigh}{neighbor}

\def\sign{\mathop{\rm sign}}
\def\ie{\emph{i.e.} }
\def\eg{\emph{e.g.} }
\def\cf{\emph{cf.} }

\title{A minimal Maxey--Riley model for the drift of \emph{Sargassum}
rafts}

\author[F.\ J.\ Beron-Vera and P.\ Miron]{F.\ J.\
Beron-Vera\thanks{Email address for correspondence:
fberon@miami.edu}\ns and P.\ Miron}

\affiliation{Department of Atmospheric Sciences, Rosenstiel School
of Marine and Atmospheric Science, University of Miami, Miami FL,
USA}

\pubyear{2020}
\volume{X}
\pagerange{X--X}
\date{Submitted 3 March 2020; revised 19 May 2020; accepted 30
July 2020.}
\begin{document}

\maketitle

\begin{abstract}
  Inertial particles (\ie with mass and of finite size) immersed
  in a fluid in motion are unable to adapt their velocities to the
  carrying flow and thus they have been the subject of much interest
  in fluid mechanics.  In this paper we consider an ocean setting
  with inertial particles elastically connected forming a network
  that floats at the interface with the atmosphere.  The network
  evolves according to a recently derived and validated Maxey--Riley
  equation for inertial particle motion in the ocean.  We rigorously
  show that, under sufficiently calm wind conditions, rotationally
  coherent quasigeostrophic vortices (which have material boundaries
  that resist outward filamentation) always possess finite-time
  attractors for elastic networks if they are anticyclonic, while
  if they are cyclonic provided that the networks are sufficiently
  stiff.  This result is supported numerically under more general
  wind conditions and, most importantly, is consistent with
  observations of rafts of pelagic \emph{Sargassum}, for which the
  elastic inertial networks represent a minimal model.  Furthermore,
  our finding provides an effective mechanism for the long range
  transport of \emph{Sargassum}, and thus for its connectivity
  between accumulation regions and remote sources.
\end{abstract}

\section{Introduction}

This paper is motivated by a desire to understand the mechanism
that leads rafts of pelagic \emph{Sargassum}---a genus of large
brown seaweed (a type of alga)---to choke coastal waters and land
on, most notably, the Caribbean Sea and beaches, phenomenon that
has been on the rise and is challenging scientists, coastal resource
managers, and administrators at local and regional levels
\citep{Langin-18}.  A raft of pelagic \emph{Sargassum} is composed
of flexible stems which are kept afloat by means of bladders filled
with gas while it drifts under the action of ocean currents and
winds (Figure \ref{fig:net}, left panel).  A mathematical model is
here conceived for this physical depiction of a drifting \emph{Sargassum}
raft as an elastic network of buoyant, finite-size or \emph{inertial}
particles that evolve according to a novel motion law
\citep{Beron-etal-19-PoF}, which has been recently shown capable
of reproducing observations \citep{Olascoaga-etal-20}.  The motion
law derives from the \emph{Maxey--Riley equation} \citep{Maxey-Riley-83},
a classical mechanics Netwon's second law that constitutes the
de-jure fluid mechanics framework for investigating inertial dynamics
\citep{Michaelides-97}.  The inability of inertial particles to
adapt their velocities to the carrying fluid flow leads to dynamics
that can be quite unlike that of fluid or Lagrangian (\ie neutrally
buoyant, infinitesimally small) particles \citep{Cartwright-etal-10}.
While largely overlooked in ``particle tracking'' in oceanography,
particularly \emph{Sargassum} raft tracking \citep{Putman-etal-18,
Johns-etal-20}, this holds true for neutrally buoyant particles,
irrespective of how small they are \citep{Babiano-etal-00,
Sapsis-Haller-10}.  The Maxey--Riley theory for inertial particle
dynamics in the ocean \citep{Beron-etal-19-PoF, Olascoaga-etal-20}
accounts for the \emph{combined effects of ocean currents and winds
on the motion of floating finite-size particles}.  Elastic interaction
among such particles unveils, as we show here, a mechanism for
long-range transport that may be at the core of connectivity of
\emph{Sargassum} between accumulation regions in the Caribbean Sea
and surroundings and possibly quite remote blooming areas in the
tropical North Atlantic from the coast of Africa \citep{Ody-etal-19}
to the Amazon River mouth \citep{Gower-etal-13}, along what has
been dubbed \citep{Wang-etal-19} the ``Great \emph{Sargassum} belt.''

\begin{figure}
  \centering%
  \includegraphics[width=1\columnwidth]{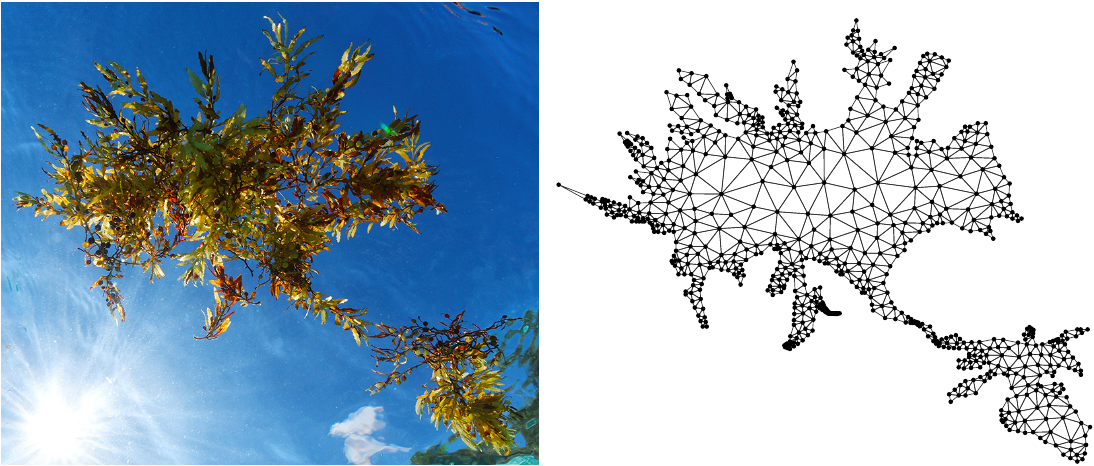}%
  \caption{(left) Floating raft of \emph{Sargassum}. Credit: Alain
  M.\ Brin, Blue Glass Photography. (right) Elastic network of finite-size,
  buoyant particles providing a minimal representation for the
  raft on the left.}
  \label{fig:net}%
\end{figure}

\section{The model}

To construct the mathematical model, we consider a (possibly
irregular) network of $N > 1$ spherical particles (beads) connected by
(massless, nonbendable) springs.  The particles are assumed to have
small radius, denoted by $a$, and to be characterized by a
water-to-particle density ratio $\delta \ge 1$ finite, so $1 -
\delta^{-1}$ approximates well \citep{Olascoaga-etal-20} reserve
volume assuming that the air-to-particle density ratio is very
small.  The elastic force (per unit mass) exerted on particle $i$,
with 2-dimensional Cartesian position $x_i = (x^1_i,x^2_i)$, by
neighboring particles at positions $\{x_j : j\in \neigh(i)\}$, is
assumed to obey Hooke's law \citep[\cf \eg][]{Goldstein-81}:
\begin{equation}
F_i = 
- \sum_{j\in \neigh(i)} k_{ij}\big(|x_{ij}| -
\ell_{ij}\big)\frac{x_{ij}}{|x_{ij}|},
\label{eq:F}
\end{equation}
$i = 1,\dotsc,N$, where 
\begin{equation}
  x_{ij} := x_i - x_j;
  \label{eq:xij}
\end{equation}
$k_{ij} \ge 0$ is the stiffness (per unit mass) of the spring
connecting particle $i$ with neighboring particle $j$; and $\ell_{ij}
\ge 0$ is the length of the latter at rest.   Elastic network models
are commonly employed to represent biological macromolecules in the
study of dynamics and function of proteins \citep{Bahar-etal-97}.
Elastic chain models, a particular form of elastic network models,
are used to represent polymers \citep{Bird-etal-77}.  A relevant
recent application \citep{Picardo-etal-18} is the investigation of
preferential sampling of inertial chains in turbulent flow.

According to the Maxey--Riley theory for inertial ocean motion
\citep{Beron-etal-19-PoF, Olascoaga-etal-20}, a particle of the
elastic network, when taken in isolation, evolves according to the
following 2nd-order ordinary differential equation (Appendix A)
\begin{equation}
  \ddot x + \left(f + \tfrac{1}{3}R\omega\right)\dot x^\perp +
  \tau^{-1} \dot x = R\frac{\D{v}}{\D{t}} + R\left(f +
  \tfrac{1}{3}\omega\right)v^\perp + \tau^{-1}u, 
  \label{eq:MR}
\end{equation}
where
\begin{equation}
  u := (1-\alpha)v + \alpha v_\mathrm{a} 
  \label{eq:u}
\end{equation}
and $\perp$ represents a $+\frac{1}{2}\pi$ rotation.
Time-and/or-position-dependent quantities in \eqref{eq:MR}--\eqref{eq:u}
are: the (horizontal) velocity of the \emph{water}, $v$, with
$\frac{\D{}}{\D{t}}v = \partial_t v + (\nabla v)v$ where $\nabla$
is the gradient operator in $\mathbb R^2$; the \emph{water}'s
vorticity, $\omega$; the \emph{air} velocity, $v_\mathrm{a}$; and
the Coriolis ``parameter,'' $f = f_0 + \beta x^2$, where $f_0 =
2\Omega\sin\vartheta_0$ and $\beta = 2a_\odot^{-1}\Omega\cos\vartheta_0$
with $\Omega$ and $a_\odot$ being Earth's angular velocity magnitude
and mean radius, respectively, and  $\vartheta_0$ being reference
latitude. Quantities independent of position and time in
\eqref{eq:MR}--\eqref{eq:u} in turn are:
\begin{equation}
  R(\delta) : = \frac{1 -
  \frac{1}{2}\Phi(\delta)}{1 - \frac{1}{6}\Phi(\delta)} \in
  [0,1);
\end{equation}
\begin{equation}
  \tau(\delta) := \frac{1-\frac{1}{6}\Phi(\delta)}{\left(1 + (1 -
  \gamma)\Psi(\delta)\right)\delta^4}\cdot \frac{a^2\rho}{3\mu}
  > 0,
\end{equation}
which measures the inertial response time of the medium to the
particle ($\rho$ is the assumed constant water density and $\mu$
the water dynamic viscosity); and 
\begin{equation}
  \alpha(\delta) := \frac{\gamma\Psi(\delta)}{1 + (1 - \gamma)\Psi(\delta)}
  \in [0,1),
\end{equation}
which makes the convex combination \eqref{eq:u} a weighted average
of water and air velocities ($\gamma \approx 1/60$ is the air-to-water
viscosity ratio).  Here 
\begin{equation}
  \Phi(\delta) := \frac{\mathrm{i}\sqrt{3}}{2}
  \left(\varphi(\delta)^{-1} - \varphi(\delta)\right) -
  \frac{1}{2}(\varphi(\delta)^{-1} + \varphi(\delta)) + 1 \in [0,2)
\end{equation}
is the fraction of emerged particle piece's height, where
\begin{equation}
  \varphi(\delta) := \sqrt[3]{\mathrm{i}\sqrt{1 - (2\delta^{-1}
  - 1)^2} + 2\delta^{-1} - 1}
\end{equation}
and 
\begin{equation}
  \Psi(\delta) := \pi^{-1}\cos^{-1}(1
  - \Phi(\delta)) - \pi^{-1}(1 - \Phi(\delta)) \smash{\sqrt{1 - (1 -
  \Phi(\delta))^2}} \in [0,1),
\end{equation}
which gives the fraction of emerged particle's projected (in the
flow direction) area.  The \emph{Sargassum raft drift model} is
obtained by adding the elastic force \eqref{eq:F} to the right-hand-side
of the Maxey--Riley set \eqref{eq:MR}.  The result is a set of $N$
2nd-order ordinary differential equations, \emph{coupled} by the
elastic term, viz.,
\begin{equation}
  \ddot x_i + \left(\left. f\right\vert_i +
  \tfrac{1}{3}R\left.\omega \right\vert_i\right)\dot x_i^\perp +
  \tau^{-1} \dot x_i = R\frac{\D{\left. v\right\vert_i}}{\D{t}} +
  R\left(\left. f\right\vert_i +
  \tfrac{1}{3}\left.\omega\right\vert_i\right)\left
  .v\right\vert_i^\perp + \tau^{-1}\left.
  u\right\vert_i + F_i,
  \label{eq:SM}
\end{equation}
$i = 1,\dotsc,N$, where $\left. \right\vert_i$ means pertaining to
particle $i$.

Now, as the radius ($a$) of the elastically interacting particles
is small by assumption, the inertial response time ($\tau \propto
a^2$) is short. We write, then, $\tau = O(\varepsilon)$ where $0<
\varepsilon \ll 1$ is a parameter that we use to measure smallness
throughout this paper.  In this case $\varepsilon$ can be interpreted
as a Stokes number \citep{Cartwright-etal-10}.  That $\tau =
O(\varepsilon)$ has an important consequence: \eqref{eq:SM} represents
a singular perturbation problem involving slow, $x_i$, and fast,
$v_i = \dot x_i$, variables.  This readily follows by rewriting
\eqref{eq:SM} as a system of 1st-order ordinary differential equations
in $(x_i,v_i)$, \ie a nonautonomous $4$-dimensional dynamical system,
which reveals that while $x_i$ changes at $O(1)$ speed, $v_i$ does
it at $O(\varepsilon^{-1})$ speed.  The geometric singular perturbation
theory of Fenichel \citep{Fenichel-79, Jones-95} extended to
nonautonomous systems \citep{Haller-Sapsis-08} was applied by
\citet{Beron-etal-19-PoF} to \eqref{eq:MR} to frame its \emph{slow
manifold}, to wit, a $(2+1)$-dimensional subset $\{(x,v_\mathrm{p},t)
: v_\mathrm{p} = u(x,t) + u_\tau(x,t) + O(\varepsilon^2)\}$ of the
$(4+1)$-dimensional phase space $(x, v_\mathrm{p}, t)$ where
\begin{equation}
  u_\tau =  \tau\left(R\frac{\D{v}}{\D{t}} + R \left(f +
  \tfrac{1}{3}\omega\right) v^\perp - \frac{\D{u}}{\D{t}} -
  \left(f + \tfrac{1}{3}R\omega\right) u^\perp\right)
  \label{eq:utau}
\end{equation}
with $\frac{\D{}}{\D{t}}u = \partial_t u + (\nabla u)u$, which
normally attracts all solutions of \eqref{eq:MR} exponentially in
time. \emph{On} the slow manifold, equation \eqref{eq:MR} reduces to a
1st-order ordinary differential equation in $x$ given by $\dot x =
v_\mathrm{p} = u + u_\tau + O(\varepsilon^2)$, which represents a
nonautonomous $2$-dimensional dynamical system.  Mathematically
more tractable than the full set \eqref{eq:MR}, this reduced set
facilitated uncovering aspects of the inertial ocean dynamics such
as the occurrence of great garbage patches in the ocean's subtropical
gyres \citep{Beron-etal-16, Beron-etal-19-PoF} and the potential
role of mesoscale eddies (vortices) as flotsam traps \citep{Beron-etal-15,
Haller-etal-16, Beron-etal-19-PoF}.  Because the elastic force
\eqref{eq:F} does not depend on velocity, the geometric singular
perturbation analysis of \eqref{eq:MR} by \citet{Beron-etal-19-PoF}
applies to \eqref{eq:SM} with the only difference that the equations
on the slow manifold are coupled by the elastic force \eqref{eq:F},
namely,
\begin{equation}
 \dot x_i = v_i = \left. u\right\vert_i +  \left.
 u_\tau\right\vert_i + \tau F_i + O(\varepsilon^2),  
 \label{eq:SMslow}
\end{equation}
$i = 1,\dotsc, N$. The slow manifold of \eqref{eq:SM} is the $(2N
+ 1)$-dimensional subset $\{(x_i,v_i,t) : v_i = u(x_i,t) + u_\tau(x_i,t)
+ \tau F_i(x_i; x_j:j\in \neigh(i)) + O(\varepsilon^2),\, i = 1,
\dotsc, N\}$ of the $(4N + 1)$-dimensional phase space $(x_i,v_i,t)$,
$i = 1, \dotsc, N$.

\begin{figure}
  \centering%
  \includegraphics[width=\columnwidth]{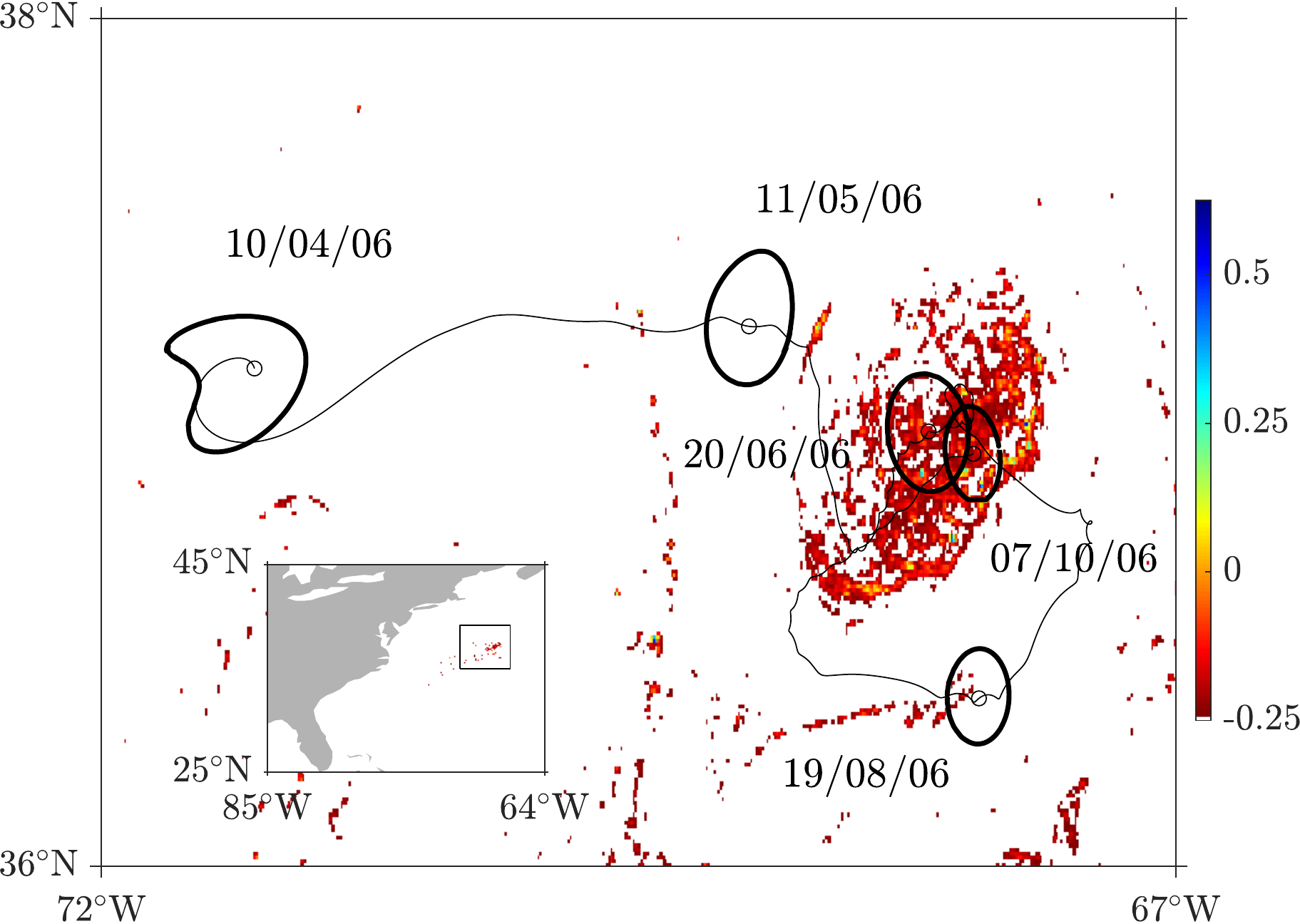}%
  \caption{Floating \emph{Sargassum} distribution inferred from
  satellite spectrometry on the first week of October 2006 in the
  region of the Northwestern Atlantic indicated in the inset map.
  \emph{Sargassum} corresponds to Maximum Chlorophyll Index (MCI)
  values exceeding $-0.25$ mW m$^{-2}$ sr$^{-1}$ nm$^{-1}$. MCI is
  inferred from the Medium Resolution Imaging Spectrometer (MERIS)
  aboard \emph{Envisat}.  Overlaid in heavy black are snapshots
  of the evolution of the boundary of a coherent material vortex
  revealed from satellite altimetry data.  The small open circle
  represents the center of the vortex and the black curve the
  corresponding trajectory.}
  \label{fig:sar}%
\end{figure}

\section{Behavior near mesoscale eddies}

Having settled on a Maxey--Riley equation for \emph{Sargassum} raft
drift, we turn to evaluate its ability to represent reality.  This
evaluation is not meant to be exhaustive; such type of evaluation
is left for a future publication. With this in mind, we consider
an actual observation of \emph{Sargassum}, in the Northwestern
Atlantic (Figure \ref{fig:sar}). This figure more precisely shows,
on the first week of October 2006, satellite-derived Maximum
Chlorophyll Index (MCI) at the ocean surface.  Floating \emph{Sargassum}
corresponds to MCI values exceeding $-0.25$ mW m$^{-2}$ sr$^{-1}$
nm$^{-1}$ \citep{Gower-etal-08, Gower-etal-13}.  Note the spiraled
shape of high-MCI distribution filling a compact region.  Overlaid
on the MCI distribution are snapshots of the evolution of a
\emph{coherent material vortex/eddy}, as extracted from satellite
altimetry measurements of sea surface height, widely used to
investigate mesoscale (50--200 km) variability in the ocean
\citep{Fu-etal-10}.  Shown in heavy black is the boundary of the
vortex; the (small) open circle and thin black curve indicate its
center and trajectory described, respectively. Below we will give
precise definitions for all these objects.  What is important to
be realized at this point is that, being material, the boundary of
such a vortex, which can be identified with the core of a cold Gulf
Stream ring (vortex) \citep{Talley-11}, cannot be traversed by
water.  Yet it may be bypassed by inertial particles, whose motion
is not tied \citep{Haller-Sapsis-08, Beron-etal-15} to \emph{Lagrangian
coherent structures} \citep{Haller-Yuan-00, Haller-16}.  But this
is not enough to explain the collection of \emph{Sargassum} inside
the ring.  In fact, this ring is cyclonic (a water particle along
the boundary circulates in the local Earth rotation's sense, which
is anticlockwise in the northern hemisphere), and inertial particles
tend to collect inside anticyclonic vortices while avoiding cyclonic
vortices, as it was formally shown by \citet{Beron-etal-19-PoF} in
agreement with a similar observed tendency of plastic debris in the
North Atlantic subtropical gyre \citep{Brach-etal-18}.  The relevant
question is whether elastic interaction alters this paradigm.

\begin{figure}
  \centering%
  \includegraphics[width=\textwidth]{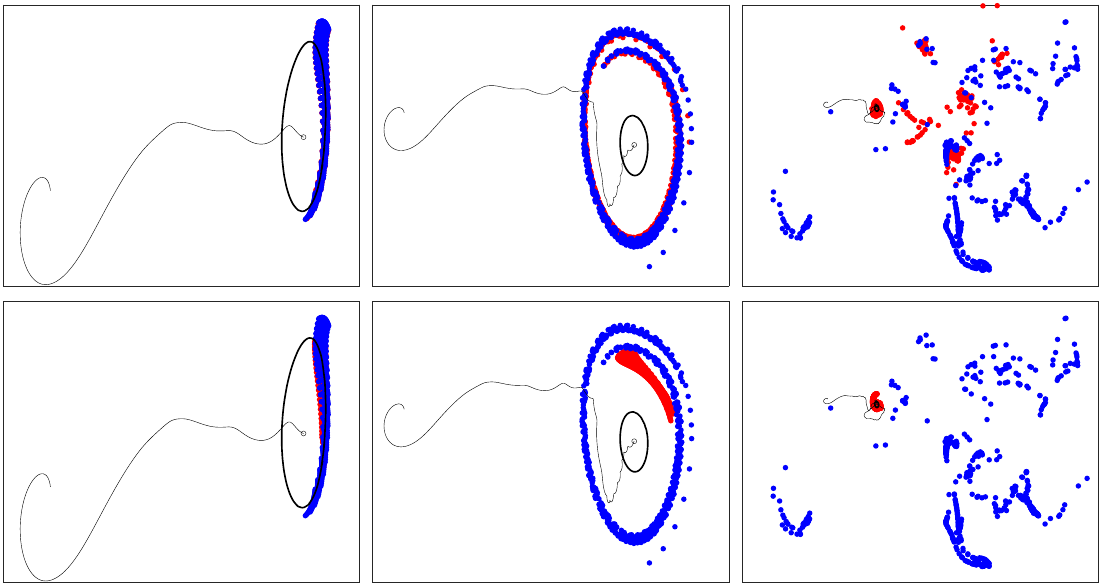}%
  \caption{Snapshots of the evolution of elastic networks of inertial
  particles (red) initially lying on the boundary of the coherent
  material vortex of Figure \ref{fig:sar}.  From left to right are
  positions 30, 60, and 180 d after initialization on 10/04/06.
  The stiffness of the network in the top panels is smaller than
  that in the bottom panels.  Blue dots, shown for reference,
  correspond to inertial particles which do not interact elastically.
  Overlaid in all panels are the boundary of the vortex (heavy
  black), center (small open circle), and corresponding trajectory
  (black curve).}
  \label{fig:pos}%
\end{figure}

\begin{figure}
  \centering%
  \includegraphics[width=\textwidth]{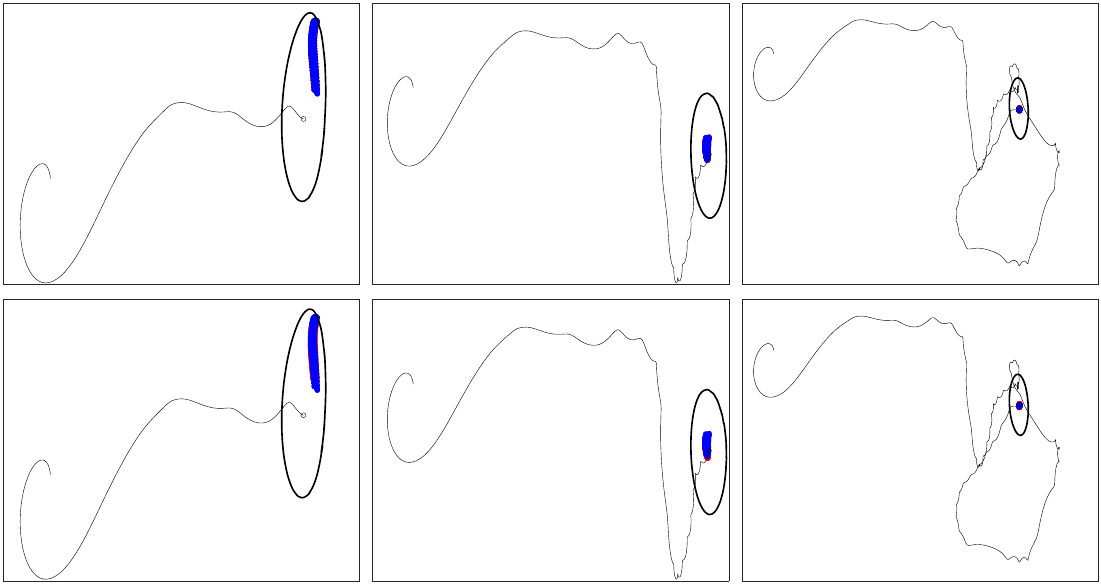}% 
  \caption{As in Fig.\ \ref{fig:pos}, but with the sign of Coriolis
  parameter ($f$) artificially set negative.}
  \label{fig:neg}%
\end{figure}

We begin by addressing this question via direct numerical
experimentation. This is done by integrating \eqref{eq:MR} for an
elastic network of inertial particles centered at a point on the
boundary of the coherent material vortex on 10/04/06.  The water
velocity $v$ is inferred using altimetry, following standard practice
\citep[\eg][]{Beron-etal-08-GRL}.  In turn, the air velocity
($v_\mathrm{a}$) is obtained from reanalysis \citep{Dee-etal-11}.
While these velocities provide an admittedly imperfect representation
of the carrying flow, they are data based and hence enable a
comparison with observed behavior. Parameters characterizing the
carrying fluid system are set to mean values, namely, $\rho = 1025$
kg\,m$^{-3}$, $\rho_\mathrm{a} = 1.2$ kg\,m$^{-3}$, $\mu = 0.001$
kg\,m$^{-1}$s$^{-1}$, and $\mu_\mathrm{a} = 1.8 \times 10^{-5}$
kg\,m$^{-1}$s$^{-1}$.  The initial network is chosen to be a square
of 12.5-km side (it could be chosen irregular, if desired, as that
one obtained from Delaunay triangulation of polygonal regions
spanning the area covered by the \emph{Sargassum} raft in Figure
\ref{fig:net}).  The network's springs are of equal length at rest,
$\ell_{ij} = 0.5$ m.  The beads, totalling $n = 625$, have a common
radius $a = 0.1$ m.  The buoyancies of the beads are all taken the
same and equal to $\delta = 1.25$, which has been found appropriate
for \emph{Sargassum} \citep{Olascoaga-etal-20}. The resulting
inertial parameters $\alpha = 5.9\times 10^{-3}$, $R = 0.6$, and
$\tau = 4.1\times 10^{-2}$ d. Shown in red in Figure \ref{fig:pos}
are snapshots (on 11/05/06, 20/06/06, and 07/10/06) of the evolution
of the network for two different stiffness values, $k_{ij} = 4.25$
d$^{-2}$ (top) and $k_{ij} = 425$ d$^{-2}$ (bottom).  For reference,
inertial particles, unconstrained by elastic forces, \ie with motion
obeying \eqref{eq:MR} or \eqref{eq:SM} with $F_i = 0$, are shown
in blue, and the boundary and trajectory of the center of the
coherent material vortex are shown in black.  The inertial particles,
consistent with Beron-Vera \emph{et al.}'s (2019)
\nocite{Beron-etal-19-PoF} prediction, are repelled away from the
vortex.  By contrast, the elastic network of inertial particles
remains close to it when $k_{ij} = 4.25$ d$^{-2}$ or, much more
consistent with the observed \emph{Sargassum} distribution in Figure
\ref{fig:sar}, collect inside the vortex when $k_{ij} = 425$ d$^{-2}$.
In Figure \ref{fig:neg} we show the results of the same numerical
experiments when the sense of the planet's rotation is artificially
changed, mimicking conditions in the southern hemisphere.  This is
achieved by multiplying the Coriolis parameter ($f$) by $-1$.  The
effect of this alteration first is a change in the polarity of the
vortex from cyclonic to anticyclonic.  The second, more important
effect is that the inertial particles of the network, irrespective
of whether they elastically interact or not, are attracted into the
vortex.  Next we show how analytic treatment of the reduced
Maxey--Riley set \eqref{eq:SMslow} sheds light on the numerically
inferred behavior just described.

\section{A formal result}

With the above goal in mind, we first make the coherent material
vortex notion precise.  This is done by considering the
\emph{Lagrangian-averaged vorticity deviation} or \emph{LAVD} field:
\cite{Haller-etal-16}
\begin{equation}
  \LAVD_{t_0}^t(x_0) := \int_{t_0}^t
  |\omega(F_{t_0}^{t'}(x_0),t'),t') - \bar{\omega}(t')|\d{t'}, 
  \label{eq:lavd}
\end{equation}
where 
\begin{equation}
  \bar{\omega}(t) = \frac{1}{\area U(t)}\int_{U(t)}
  \omega(x,t)\,\mathrm{d}^2x,
  \label{eq:bar}
\end{equation}
which is an average of the vorticity over a region of water $U(t)
= F_{t_0}^tU(t_0)$. Here $F_{t_0}^t$ is the flow map that takes
water particle positions $x_0$ at time $t_0$ to positions $x$ at
time $t$. As defined by \citet{Haller-etal-16}, a \emph{rotationally
coherent vortex} over  $t \in [t_0, t_0 + T]$ is an evolving material
(water) region $V(t)\subset U(t)$, $t \in [t_0, t_0 + T]$, such
that its time-$t_0$ position is enclosed by the outermost, sufficiently
convex isoline of $\LAVD_{t_0}^{t_0+T}(x_0)$ around a local
(nondegenerate) maximum (resp., minimum) is $T > 0$ (resp., $T <
0$).  (To be more precise, a region $V(t)$ may contain several local
extrema \citep{Beron-etal-19-PNAS}, but we conveniently exclude
from consideration such situations here to enable a straightforward
definition of vortex center \citep{Haller-etal-16}.) As a consequence,
the elements of the boundaries of such material regions $V(t)$
complete the same total material rotation relative to the mean
material rotation of the whole water mass in the domain $U(t)$ that
contains them.  This property of the boundaries tends
\citep{Haller-etal-16} to restrict their filamentation to be mainly
tangential under advection from $t_0$ to $t_0 + T$.  Furthermore,
the ensuing water-holding-property of rotationally coherent
eddies and related elliptic Lagrangian coherent structures
\citep{Haller-Beron-13, Haller-Beron-14, Farazmand-Haller-16,
Haller-etal-18}, verified numerically extensively \citep{Haller-etal-16,
Beron-etal-19-PNAS} and observed in controlled laboratory experiments
\citep{Tel-etal-18, Tel-etal-20} and field surveys involving in-situ
(buoy trajectories) and remote (satellite-inferred chlorophyll
distributions) measurements \citep{Beron-etal-18}, can be so enduring
\citep{Wang-etal-15, Wang-etal-16} for the water-holding-property
to provide a very effective long-range transport mechanism in the
ocean consistent with traditional oceanographic expectation
\citep{Gordon-86}.   The material vortex in Figure \ref{fig:sar}
(and also \ref{fig:pos} and \ref{fig:neg}) is of the rotationally
coherent class just described.  It was obtained by applying LAVD
analysis on $t_0 =$ 07/10/06, a day of the week when the \emph{Sargassum}
raft observation in Figure \ref{fig:sar} was acquired, using $T =
-180$ d.  This turned out to be the longest backward-time integration
from which a closed LAVD isoline with a stringent convexity deficiency
of $10^{-3}$ was possible to find.  It represents a rather long
backward-time integration, which dates the ``genesis'' of the
rotationally coherent vortex around $t_0 + T =$ 10/04/06.  Figure
\ref{fig:sar} not only shows the vortex boundary on detection date
($t_0$), but also several advected images of it under the backward-time
flow out to $t_0 + T$.

The second step in reaching the goal above is to assume that set
\eqref{eq:SMslow}, which attracts all solutions of \eqref{eq:SM},
can be approximated by
\begin{equation}
 \dot x_i = v_i = gf_0^{-1}\nabla^\perp\left.\eta\right\vert_i +
 \tau \big(g(1-\alpha-R)\nabla\left.\eta\right\vert_i + F_i\big)
 \label{eq:SMslow-qg}
 \end{equation}
$+\,O(\varepsilon^2)$, $i = 1,\dotsc, N$, which is justified as
follows.  First, the near surface ocean flow is in quasigeostrophic
balance \citep{Pedlosky-87}, as it can be expected for mesoscale
ocean flow \citep{Fu-etal-10}. Interpreting $\varepsilon$ as a
Rossby number \citep{Pedlosky-87}, this means that $v =
gf_0^{-1}\nabla^\perp\eta + O(\varepsilon^2)$, where $g$ is gravity
and $\eta = O(\varepsilon)$ sea surface height, $\partial_t =
O(\varepsilon)$, and $f = f_0 + O(\varepsilon)$.  Second,  the
elastic interaction does not alter the nature of the critical and
slow manifolds, which is guaranteed by making $F_i = O(\varepsilon)$.
Third, $\alpha = O(\varepsilon)$, at least, consistent with it being
very small (a few percent) over a large range of buoyancy ($\delta$)
values; \cf Figure 2 of \citet{Beron-etal-19-PoF}.  Indeed, taking
$\delta = 1.25$ (as it was found appropriate by \citet{Olascoaga-etal-20}
for \emph{Sargassum}), recall we estimated $\alpha \approx 5\times
10^{-3}$.  This is actually quite small, and more consistent with
$\alpha = O(\varepsilon^3)$ for a Rossby number that typically
characterizes mesoscale flow ($\varepsilon = 0.1$).  Note that this
makes $\alpha v_\mathrm{a} = O(\varepsilon^3)$ for an $O(1)$ near
surface atmospheric flow.  But this would not be consistent with
the quasigeostrophic ocean flow assumption. So we require, fourth,
that $v_\mathrm{a} = O(\varepsilon^2)$, at least, \ie the wind field
over the period of interest is sufficiently weak (calm).

\begin{figure}
  \centering%
  \includegraphics[width=.75\textwidth]{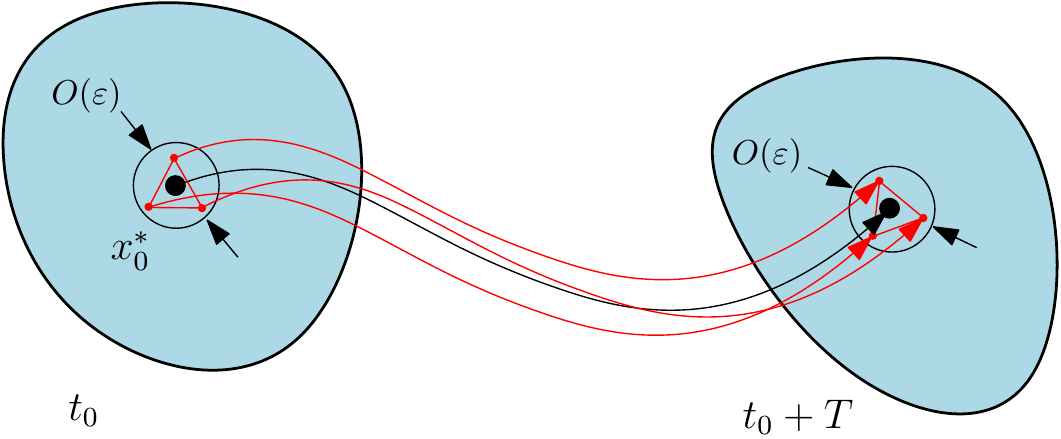}% 
  \caption{By smooth dependence of the solutions of \eqref{eq:SMslow-qg}
  on parameters, an elastic network of inertial particles initially
  $O(\varepsilon)$-close to the center of $x_0^*$ of a rotationally
  coherent vortex will remain $O(\varepsilon)$-close to the trajectory
  flowing from it over a finite-time interval $[t_0, t_0+T]$.}
  \label{fig:eps}%
\end{figure}

Now, let $\mathbf x := (x_1^1, \dotsc, x_N^1, x_1^2, \dotsc, x_N^2)$
and $\mathbf v := (v_1^1, \dotsc, v_N^1, v_1^2, \dotsc, v_N^2)$.
Then write \eqref{eq:SMslow-qg} as
\begin{equation}
  \dot{\mathbf x} = \mathbf v(\mathbf x,t)
  \label{eq:dxdt}
\end{equation}
We denote $\mathbf F_{t_0}^t$ the corresponding flow map, namely,
$\mathbf F_{t_0}^t(\mathbf x_0) := \mathbf x(t; \mathbf x_0, t_0)$
where $\mathbf x_0 = \mathbf x(t_0)$. Following \citet{Haller-etal-16}
closely, we invoke Liouville's theorem \citep[\eg][]{Arnold-89} and
note that a trajectory $\smash{\mathbf F_{t_0}^t}(\mathbf x_0)$ is
\emph{overall forward attracting over} $t\in [t_0,t_0+T]$, $T>0$
(resp., $T<0$), if $\det\smash{\bD{\mathbf F}_{t_0}^{t_0+T}}(\mathbf
x_0) < 1$ (resp., $\det\smash{\bD{\mathbf F}_{t_0}^{t_0+T}}(\mathbf
x_0) > 1$).  Let us suppose now that the time-$t_0$ position of the
network of elastically connected inertial particles is very close
to the center of a rotationally coherent vortex, given by
\citep{Haller-etal-16}
\begin{equation}
  x_0^* = \argmax_{x_0\in V(t_0)}
  \LAVD_{t_0}^{t_0+T}(x_0),
  \label{eq:x0*}
\end{equation}
which is expected to exist in a well-defined fashion when the ocean
flow is quasigeostrophic, as we have assumed here.  We write the
above formally as $|x_i(t_0) - x_0^*| = O(\varepsilon)$, $i =
1,\dotsc,N$. Then, by smooth dependence of the solutions of
\eqref{eq:SMslow-qg} on parameters, for $t\in [t_0, t_0+T]$ finite,
one has
\begin{equation}
  x_i(t;x_i(t_0),t_0) = F_{t_0}^t(x_0^*)
  + O(\varepsilon), 
  \label{eq:xi}
\end{equation}
$i = 1,\dotsc,N$, where $F_{t_0}^t$ is the flow map generated by
the quasigeostrophic ocean velocity field $gf_0^{-1}\nabla^\perp\eta$
(Figure \ref{fig:eps}).  With this in mind, we find (Appendix B)
\begin{equation}
  \det\bD{\mathbf F}_{t_0}^{t_0+T}(\mathbf x_0) = \exp
  \tau(\mathcal A+\mathcal B)
  \label{eq:detdf}
\end{equation}
$+\,O(\varepsilon^2)$, where
\begin{equation}
  \mathcal A := gN(1 - \alpha -R)\sign_{t\in
  [t_0,t_0+T]}\big(T\nabla^2\eta(F_{t_0}^t(x_0^*),t)\big)\cdot
  \left\vert\int_{t_0}^{t_0+T}
  |\nabla^2\eta(F_{t_0}^t(x_0^*),t)|\d{t}\right\vert
  \label{eq:detdf-A}
\end{equation}
and
\begin{equation}
   \mathcal B : = -T\sum_{i=1}^N\sum_{j\in \neigh(i)}k_{ij}.
	 \label{eq:detdf-B}
 \end{equation}
Noting that $1-\alpha-R \ge 0$, it finally follows that:
\begin{thm} 
 $F_{t_0}^{t}(x_0^*)$ is locally forward attracting overall over $t\in
[t_0,t_0+T]$:
\begin{enumerate}
  \item for all $k_{ij}$ when $\sign_{t\in
  [t_0,t_0+T]}\nabla^2\eta(F_{t_0}^t(x_0^*),t) < 0$; and
  \item provided that
   \begin{equation}
     |T|\sum_{i=1}^N\sum_{j\in \neigh(i)} k_{ij} >
	  gN(1-\alpha-R)
	  \left\vert\int_{t_0}^{t_0+T}
     |\nabla^2\eta(F_{t_0}^t(x_0^*),t)|\d{t}\right\vert 
     \label{eq:k}
   \end{equation}
   when $\sign_{t\in
  [t_0,t_0+T]}\nabla^2\eta(F_{t_0}^t(x_0^*),t) > 0$.
\end{enumerate}
\end{thm}

Since $\omega = gf_0^{-1}\nabla^2\eta + O(\varepsilon^2)$, the above
result says that \emph{the center of a cyclonic rotationally coherent
quasigeostrophic eddy represents a finite-time attractor for elastic
networks of inertial particles in the presence of calm winds if
they are sufficiently stiff}, \emph{while that of an anticyclonic
eddy irrespective of how stiff}.  The minimal stiffness $k_{\min}$
required for a cyclonic eddy center to attract an elastic inertial
network over finite time decreases with network's size.  This can
be readily seen assuming that the stiffness is the same for all
pairs of elastically connected particles, say, $k_{ij} = k$, and
considering a square network with $N = n^2$ elements.  In such a
case one easily computes $\smash{\sum_{i=1}^N\sum_{j\in \neigh(i)}}
= 4n (n - 1)$ and thus $k_{\min} = \smash{\frac{n}{4(n - 1)}}(1-\alpha-R)
|f_0T^{-1}\LAVD_{t_0}^{t_0+T}(x_0^*)| + O(\varepsilon^2)$, which
decays to a value bounded away from 0 as $n\to\infty$. (In getting
the last result we have relied on the fact that $U(t)$ in \eqref{eq:bar}
can be taken as large as desired, \eg $\area U(t) = O(\varepsilon^{-1})$
as we have specifically set.) So as the size of the network increases,
the condition on the stiffness is expected to be more easily
satisfied.  Similarly, this condition is easier to be fulfilled as
the buoyancy of the particles approaches neutrality; indeed,
$\lim_{\delta\to 1}k_{\min} = 0$. Note, on the other hand, that
$\lim_{n\to 1} k_{\min} = \infty$.  Thus, as expected, the result
of \citet{Beron-etal-19-PoF} for isolated inertial particles is
recovered: while anticyclonic eddy centers attract finite-size
particles floating at the ocean surface, cyclonic ones always repel
them away.  It is important to realize that statements on the
existence of finite-time attractors inside rotationally coherent
eddies do not say anything about basins of attraction.  Yet the
expectation, verified numerically above in qualitative agreement
with remote-sensing data, is that mesoscale eddies will in general
trap \emph{Sargassum} rafts if they initially lie near their
boundaries (the sensitivity analysis in Appendix C provides further
numerical support for this expectation).

\section{Concluding remarks}

The above formal result provides an explanation for the behavior
of the elastic network in Figures \ref{fig:pos} and \ref{fig:neg}.
This encourages us to speculate that \emph{Sargassum} rafts should
behave similarly.  Of course, there are additional (physical)
processes in the ocean that may also play a role.  For instance,
downwelling associated with submesoscale (less than 10 km) motions
can lead to surface convergence of flotsam.  While such convergence
has been recently observed \citep{DAsaro-etal-18}, numerical
simulations and theoretical arguments \citep{McWilliams-16} suggest
that this should happen at the periphery of submesoscale cyclonic
vortices, where density contrast is large.  Yet, consistent with
this work, initial inspection of satellite images is revealing
\citep{Trinanes-20} that \emph{Sargassum} collection is not restricted
to vortex peripheries and further that both cyclonic and anticyclonic
eddies trap \emph{Sargassum}.

We note too that pelagic \emph{Sargassum} is reportedly
\citep{Sheinbaum-20} observed to sometimes be found beneath the sea
surface, which can be a result of downwellings and/or reductions
of the buoyancy of the rafts as they absorb water or undergo
physiological transformations.  The effects of the latter can be
incorporated into the minimal model of this paper, partially at
least, by making $\delta \ge 1$ a function of time, as it has been
done previously \citep{Tanga-Provenzale-94} in the standard
Maxey--Riley set.  Full representation, beyond the scope at present,
of possible three-dimensional aspects of the motion of \emph{Sargassum}
rafts will require one to consider the (vertical) buoyancy force
along with a reliable representation of the three components of the
ocean velocity field, coupled with an ecological model of
\emph{Sargassum} life cycle.

We close by noting that satellite-altimetry observations reveal a
dominant tendency of mesoscale eddies of either polarity to propagate
westward \citep{Morrow-etal-04a, Chelton-etal-11a} consistent with
theoretical argumentation \citep{Nof-81a, Cushman-etal-90, Graef-98,
Ripa-JPO-00b}.  This observational evidence, along with the additional
observational evidence on the long-range transport capacity of
eddies \citep{Wang-etal-15, Wang-etal-16, Beron-etal-18}, makes the
result of this paper a potentially very effective mechanism for the
connectivity of \emph{Sargassum} between the Caribbean Sea and
remote regions in the tropical North Atlantic.  Clearly, a comprehensive
modeling effort is needed to verify this hypothesis.  The are several
parameters that will require specification, which may be obtained
from a study of the architecture of \emph{Sargassum} rafts or,
alternatively, from observed evolution (as inferred from satellite
imagery) via regression or learning \citep[\eg][]{Aksamit-etal-20}.

\acknowledgements

The authors report no conflict of interest. The altimeter products
are produced by SSALTO/DUCAS and distributed by AVISO with support
from CNES (http://\allowbreak www.\allowbreak aviso.\allowbreak
oceanobs).  The ERA-Interim reanalysis is produced by ECMWF and is
available from http:/\allowbreak /www.\allowbreak ecmwf.\allowbreak
int.  MERIS satellite images are provided by ESA through the G-Pod
online platform (https://\allowbreak gpod.\allowbreak eo.\allowbreak
esa.\allowbreak int).

\appendix

\section{Review of the Maxey--Riley set \eqref{eq:MR}}

The exact motion of \emph{inertial particles} obeys the Navier--Stokes
equation with moving boundaries as such particles are extended
objects in the fluid with their own boundaries.  This results in
complicated partial differential equations which are hard to solve
and analyze.  Here, as well as in \citet{Beron-etal-19-PoF}, the
interest is in the approximation, formulated in terms of an ordinary
differential equation, provided by the \emph{Maxey--Riley equation}
\citep{Maxey-Riley-83}, the de-jure fluid mechanics paradigm for
inertial particle dynamics.

Such an equation is a classical mechanics Newton's second law with
several forcing terms that describe the motion of solid spherical
particles immersed in the unsteady nonuniform flow of a homogeneous
viscous fluid.  Normalized by particle mass, $m_\mathrm{p} =
\frac{4}{3}\pi a^3\rho_\mathrm{p}$, the relevant forcing terms for
the \emph{horizontal} motion of a sufficiently small particle,
excluding so-called Faxen corrections and the Basset-Boussinesq
history or memory term, are: 1) the \emph{flow force} exerted on
the particle by the undisturbed fluid,
\begin{equation}
  F_\mathrm{flow} =
  \frac{m_\mathrm{f}}{m_\mathrm{p}}\frac{\D{v_\mathrm{f}}}{\D{t}} 
  \label{eq:FF},
\end{equation}
where $m_\mathrm{f} = \frac{4}{3}\pi a^3\rho_\mathrm{f}$ is the
mass of the displaced fluid (of density $\rho_\mathrm{f}$), and
$\smash{\frac{\D{v_\mathrm{f}}}{\D{t}}}$ is the material derivative
of the fluid velocity ($v_\mathrm{f}$) or its total derivative taken
along the trajectory of a fluid particle, $x = X_\mathrm{f}(t)$,
i.e., $\smash{\frac{\D{v_\mathrm{f}}}{\D{t}}} =
\smash{\left[\frac{\d{}}{\d{t}}v_\mathrm{f}(x,t)\right]_{x=X_\mathrm{f}(t)}}
= \partial_t v_\mathrm{f} + (\nabla v_\mathrm{f}) v_\mathrm{f}$;
3) the \emph{added mass force} resulting from part of the fluid
moving with the particle,
\begin{equation}
  F_\mathrm{mass} =
  \frac{\frac{1}{2}m_\mathrm{f}}{m_\mathrm{p}}\left(\frac{\D{v_\mathrm{f}}}{\D{t}}
  - \dot v_\mathrm{p}\right)
  \label{eq:AM},
\end{equation}
where $\dot v_\mathrm{p}$ is the acceleration of an inertial particle
with trajectory $x = X_\mathrm{p}(t)$, i.e., $\dot v_\mathrm{p} =
\smash{\frac{\d{}}{\d{t}}\left[v_\mathrm{p}(x,t)\right]_{x=X_\mathrm{p}(t)}}
= \partial_t v_\mathrm{p}$ where $v_\mathrm{p} = \partial_t
X_\mathrm{p} = \dot x$ is the inertial particle velocity; 2) the
\emph{lift force}, which arises when the particle rotates as it
moves in a (horizontally) sheared flow,
\begin{equation}
  F_\mathrm{lift} =
  \frac{\frac{1}{2}m_\mathrm{f}}{m_\mathrm{p}}\omega_\mathrm{f}
  (v_\mathrm{f} - v_\mathrm{p})^\perp,
  \label{eq:FL}
\end{equation}
where $\omega_\mathrm{f} = \partial_1 v^2_\mathrm{f} - \partial_2
v^1_\mathrm{f}$ is the (vertical) vorticity of the fluid; and 4)
the \emph{drag force} caused by the fluid viscosity,
\begin{equation}
  F_\mathrm{drag} = \frac{12\mu_\mathrm{f}
  \frac{A_\mathrm{f}}{\ell_\mathrm{f}}}{m_\mathrm{p}}
  (v_\mathrm{f} - v_\mathrm{p}),
  \label{eq:SD}
\end{equation}
where $\mu_\mathrm{f}$ is the dynamic viscosity of the fluid, and
$A_\mathrm{f}$ ($=\pi a^2$) is the projected area of the particle
and $\ell_\mathrm{f}$ ($=2a$) is the characteristic projected length,
which we have intentionally left unspecified for future appropriate
evaluation. 

The above forces are included in the original formulation by
\citet{Maxey-Riley-83}, except for the lift force \eqref{eq:FL},
due to \citet{Auton-87} and a form of the added mass term different
than \eqref{eq:AM}, which corresponds to the correction due to
\citet{Auton-etal-88}. The specific form of lift force \eqref{eq:FL}
can be found in \citet[Chapter 4]{Montabone-02} \citep[\cf similar
forms in ][]{Henderson-etal-07, Sapsis-etal-11}.

To derive equation \eqref{eq:MR}, \citet{Beron-etal-19-PoF} first accounted
for the geophysical nature of the fluid by including the Coriolis
force.  (In an earlier geophysical adaptation of the Maxey--Riley
equation \citep{Provenzale-99}, the centrifugal force was included
as well, but this is actually balanced out by the gravitational
force on the horizontal plane.) This amounts to replacing \eqref{eq:FF}
and \eqref{eq:AM} with
\begin{equation}
  F_\mathrm{flow} =
  \frac{m_\mathrm{f}}{m_\mathrm{p}}\left(\frac{\D{v_\mathrm{f}}}{\D{t}}
  + f v_\mathrm{f}^\bot\right) 
\end{equation}
and
\begin{equation}
  F_\mathrm{mass} =
  \frac{\frac{1}{2}m_\mathrm{f}}{m_\mathrm{p}}\left(\frac{\D{v_\mathrm{f}}}{\D{t}}
  + f v_\mathrm{f}^\bot - \dot v_\mathrm{p} - f
  v_\mathrm{p}^\bot\right),
\end{equation}
respectively. 

Then, noting that fluid variables and parameters take different
values when pertaining to seawater or air, \eg
\begin{equation}
 v_\mathrm{f}(x,z,t) = 
  \begin{cases} 
	v_\mathrm{a}(x,t) & \text{if } z \in (0,h_\mathrm{a}], \\
   v(x,t)  & \text{if } z \in [-h,0),
  \end{cases}
\end{equation}
\citet{Beron-etal-19-PoF} wrote
\begin{equation}
  \dot v_\mathrm{p} + f v_\mathrm{p}^\perp = \langle F_\mathrm{flow}\rangle
  + \langle F_\mathrm{mass}\rangle + \langle F_\mathrm{lift}\rangle
  + \langle F_\mathrm{drag}\rangle, 
  \label{eq:mr}
\end{equation}
where $\langle\,\rangle$ is an average over $z\in [-h,h_\mathrm{a}]$.
After some algebraic manipulation, equation \eqref{eq:MR} follows
upon making $\ell = \ell_\mathrm{a} = \delta^{-3}h$ as suggested
by observations \citep{Olascoaga-etal-20}, and assuming $\delta_\mathrm{a}
\ll 1$ with the static stability considerations in \S IV.B of
\citet{Olascoaga-etal-20} in mind.

\section{Derivation of equations
\eqref{eq:detdf}--\eqref{eq:detdf-B}}

We begin by decomposing the elastic force as $F_i = A_i + B_i$,
where
\begin{equation}
  A_i := - \sum_{j\in \neigh(i)} k_{ij}x_{ij},\quad
  B_i :=   \sum_{j\in \neigh(i)}
  k_{ij}\ell
  _{ij}\frac{x_{ij}}{|x_{ij}|}.
  \label{eq:AB}
\end{equation}
Then we note
\begin{align}
  \nabla_iA_i = -\sum_{j\in \neigh(i)}k_{ij}\nabla_ix_{ij} = -
  \sum_{j\in \neigh(i)}k_{ij}\Id^{2\times 2}
\end{align}
and thus
\begin{align}
  \sum_{i=1}^N\trace\nabla_iA_i =
  -\sum_{i=1}^N\sum_{j\in \neigh(i)}k_{ij}\trace\Id^{2\times 2} = -2
  \sum_{i=1}^N\sum_{j\in \neigh(i)}k_{ij}.
  \label{eq:traceA}
\end{align}
Now, let $x,y\in \mathbb R^2$ and note
\begin{align}
  \nabla_x\frac{x-y}{|x-y|} 
  & = 
  \frac{\nabla_xx}{|x-y|} - \frac{(x-y)2(x-y)^\top\nabla_xx}{2|x-y|^3}\nonumber\\ 
  &= 
  \frac{\Id^{2\times 2}}{|x-y|} -
  \frac{(x-y)(x-y)^\top\Id^{2\times 2}}{|x-y|^3}\nonumber\\
  &= 
  \frac{\Id^{2\times 2}}{|x-y|} -
  \frac{(x-y)(x-y)^\top}{|x-y|^3}.
  \label{eq:nabla}
\end{align}
Consequently,
\begin{align}
  \trace\nabla_x\frac{x-y}{|x-y|}
  &=
  \trace
  \begin{pmatrix}
	 \frac{1}{|x-y|} & 0\\
	 0 & \frac{1}{|x-y|}
  \end{pmatrix}
  -
  \trace
  \begin{pmatrix}
	 \frac{(x_1-y_1)^2}{|x-y|^3} & \frac{(x_1-y_1)(x_2-y_2)}{|x-y|^3}\\
	 \frac{(x_2-y_2)(x_1-y_1)}{|x-y|^3} & \frac{(x_2-y_2)^2}{|x-y|^3}
  \end{pmatrix}\nonumber\\
  &= \frac{2}{|x-y|} - \frac{|x-y|^2}{|x-y|^3}\nonumber\\
  &= \frac{1}{|x-y|}.
  \label{eq:trace}
\end{align}
Using \eqref{eq:nabla} we obtain
\begin{align*}
  \nabla_iB_i 
  &= 
  -\sum_{j\in
  \neigh(i)} k_{ij}\ell_{ij}\left(-\frac{\nabla_ix_{ij}}{|x_{ij}|} +
  \frac{x_{ij}x_{ij}^\top}{|x_{ij}|^3}\right)\nonumber\\
  &= 
  -\sum_{j\in
  \neigh(i)} k_{ij}\ell_{ij}\left(-\frac{\Id^{2\times 2}}{|x_{ij}|} +
  \frac{x_{ij}x_{ij}^\top}{|x_{ij}|^3}\right),
\end{align*}
while using \eqref{eq:trace},
\begin{align}
  \sum_{i=1}^N\trace\nabla_iB_i 
  &=
  -\sum_{i=1}^N\sum_{j\in \neigh(i)}
  k_{ij}\ell_{ij} \trace \left(-\frac{\Id^{2\times 2}}{|x_{ij}|} +
  \frac{x_{ij}x_{ij}^\top}{|x_{ij}|^3}\right)\nonumber\\
  &=
  -\sum_{i=1}^N\sum_{j\in \neigh(i)} k_{ij}\ell_{ij}\left(-\frac{2}{|x_{ij}|} +
  \frac{1}{|x_{ij}|}\right)\nonumber\\
  &=
  \sum_{i=1}^N\sum_{j\in \neigh(i)}
  k_{ij}\frac{\ell_{ij}}{|x_{ij}|}.
  \label{eq:traceB}
\end{align}
Combining \eqref{eq:traceA} and \eqref{eq:traceB} we obtain
\begin{equation}
  \sum_{i=1}^N\trace\nabla_iF_i = \sum_{i=1}^N\sum_{j\in
  \neigh(i)} k_{ij} \left(\frac{\ell_{ij}}{|x_{ij}|} -2\right).
  \label{eq:traceF}
\end{equation}

Now, taking into account \eqref{eq:traceF}, from
\eqref{eq:SMslow-qg} it follows that
\begin{align}
  \trace\boldsymbol\nabla\mathbf v(\mathbf F_{t_0}^{t}(\mathbf
  x_0),t) ={}& \sum_1^N\trace\nabla_i v_i(x_i(t;x_i(t_0),t_0),t)\nonumber\\
  ={}& \tau R_\alpha
  g\sum_1^N\nabla^2\eta(x_i(t;x_i(t_0),t_0),t)\nonumber\\
  &+ \tau\sum_{i=1}^N\sum_{j\in \neigh(i)} k_{ij}
  \left(\frac{\ell_{ij}}{|x_{ij}(t;x_{ij}(t_0),t_0)|} -2\right) +
  O(\varepsilon^2)\nonumber\\ 
  ={}& \tau R_\alpha
  gN\nabla^2\eta(F_{t_0}^t(x_0^*),t) - \tau\sum_{i=1}^N\sum_{j\in
  \neigh(i)} k_{ij} + O(\varepsilon^2),
\end{align}
where $R_\alpha := 1-\alpha-R$. Here we have had \eqref{eq:xi} in mind, and
consistent with this have assumed $\ell_{ij} = O(\varepsilon)$, so
$\ell_{ij}/|x_{ij}| \sim 1$ as $\varepsilon\to 0$. Result
\eqref{eq:detdf}--\eqref{eq:detdf-B} follows upon noting that
$\det\bD{\mathbf F}_{t_0}^{t}(\mathbf x_0) =
\int_{t_0}^{t}\trace\boldsymbol\nabla \mathbf v(\mathbf
F_{t_0}^{t'}(\mathbf x_0),t')\d{t'}$, where $\boldsymbol\nabla$ is
the gradient operator in $\mathbb R^{2N}$.  $\square$

\section{Sensitivity analysis}

\begin{figure}
  \centering%
  \includegraphics[width=\textwidth]{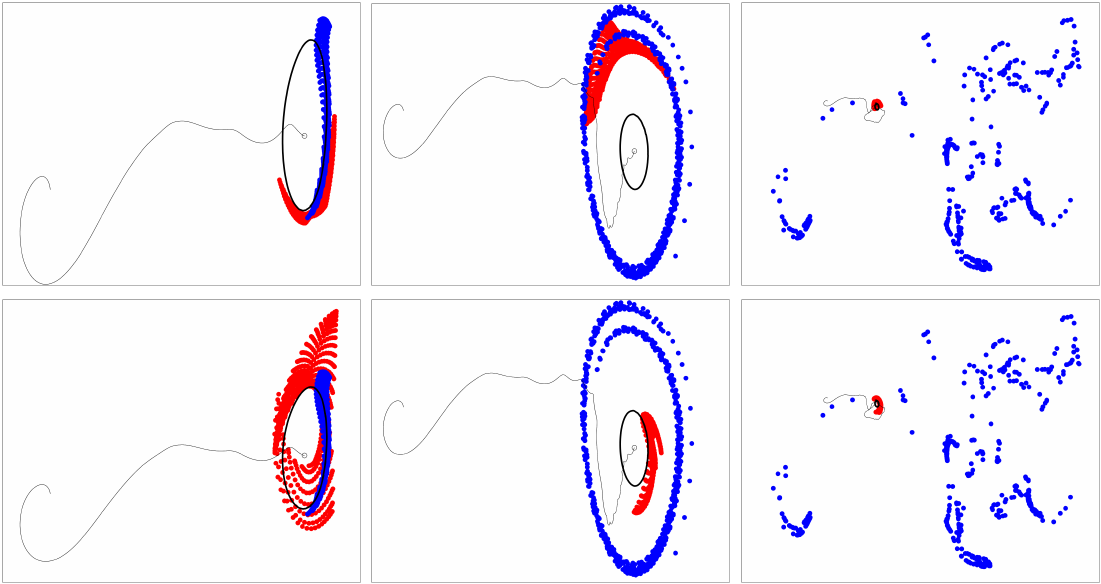}%
  \caption{As in the bottom panel of Figure \ref{fig:pos}, except
  that the initialization of the elastic networks takes place away
  from the boundary of the vortex,  at a distance equal to one (top)
  and two times as large as (bottom) the mean radius of the vortex
  (about 25 km) where the initial network's bead locations are a
  small random perturbation of the original locations forming a
  rectangular grid.}
  \label{fig:dist}%
\end{figure}

We provide further numerical support for the expectation that
mesoscales eddies should in general trap \emph{Sargassum} rafts
through a sensitivity analysis with respect to the elastic network's
initial position relative to the vortex and also the configuration
of the initial network.  This is given in Figure \ref{fig:dist},
which uses the same parameters as in bottom panels of Figure
\ref{fig:pos} except that initialization is made 25 (top) and 50
(bottom) km away from the boundary of the vortex.  These distances
correspond to about one and two times the mean radius of the vortex,
respectively.  The initial network's shape is irregular, obtained
by applying a small random perturbation on the original square
network's bead locations (Figure \ref{fig:net}).  Note the influence
of the vortex on the network. 

\begin{figure}
  \centering%
  \includegraphics[width=.5\textwidth]{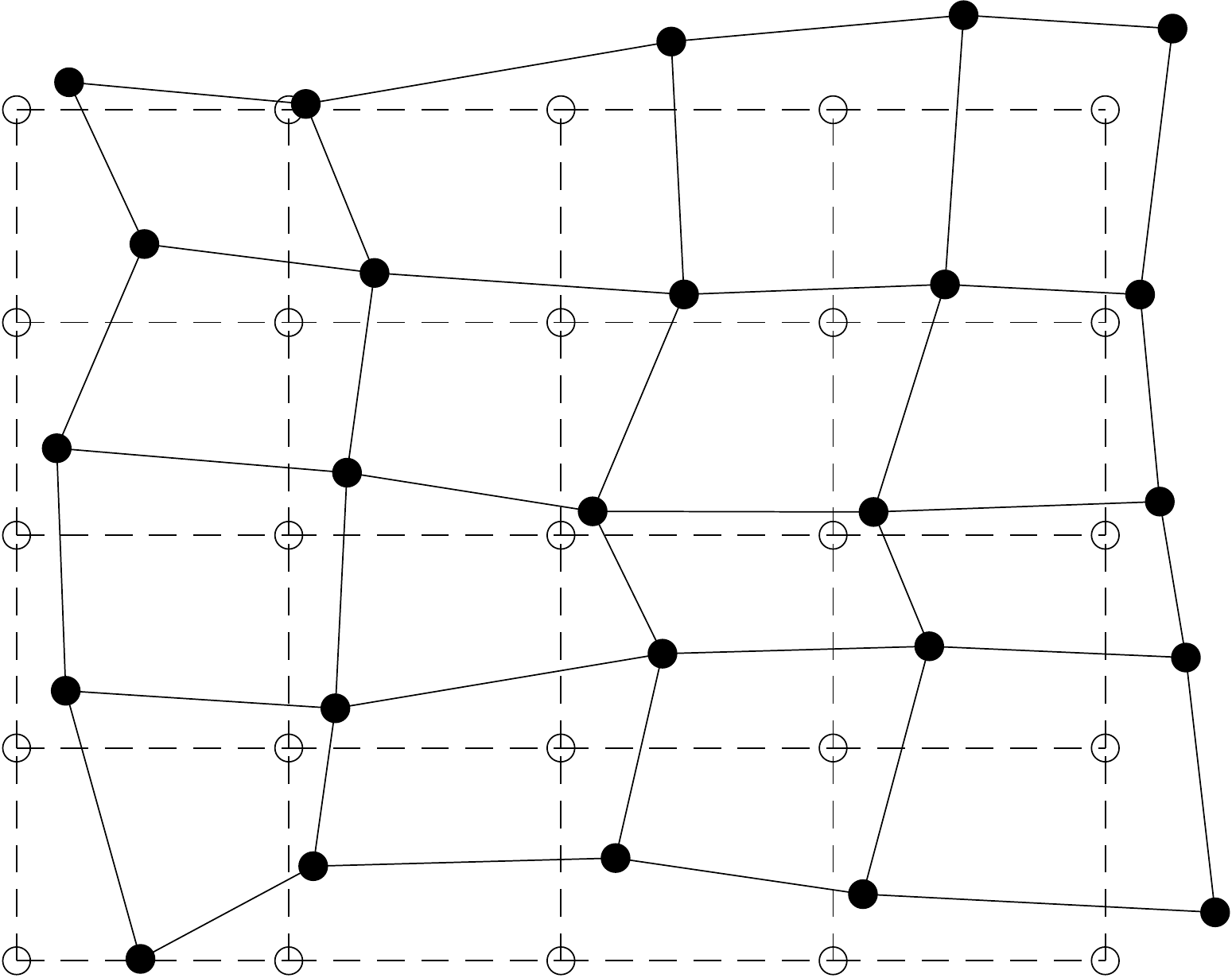}%
  \caption{Portions (of about 2.5-km side) of the initial square
  network employed in Figures \ref{fig:pos}--\ref{fig:neg} (beads
  depicted open) and the irregular network used in Figure \ref{fig:dist}
  (beads depicted solid).}
  \label{fig:net}%
\end{figure}

\bibliographystyle{jfm}
%\bibliography{fot}

\end{document}